# Electric Field Driven Magnetic Domain Wall Motion
# in Iron Garnet Film

## Nikolaev A. [1], Pyatakov A. [1,2*], Nikolaeva E. [1], Meshkov G. [1], Logginov A. [1], and Zvezdin A. [2]

[1] Physics Department, M.V. Lomonosov MSU, Leninskie gori, Moscow, 119992, Russia

[2] A. M. Prokhorov General Physics Institute, 38, Vavilova st., Moscow, 119991, Russia

* pyatakov@phys.msu.ru

**Abstract.** The dynamic observation of domain wall motion induced by electric field in magnetoelectric iron garnet film is reported. Measurements in 800 kV/cm electric field pulses gave the domain wall velocity ~ 45 m/s. Similar velocity was achieved in magnetic field pulse about 50 Oe. Reversible and irreversible micromagnetic structure transformation is demonstrated. These effects are promising for applications in spintronics and magnetic memory.

## Introduction

The conventional means of data writing based on generation of magnetic field with electric current put the limit for increasing storage density in hard disc drivers and magnetic random access memory (MRAM) [1]. The inductive coils and conducting lines that are used to generate magnetic field suffer from energy losses, and at further downscaling from electromigration, i.e. the directed motion of ions as a result of current flowing, that causes the progressive damage of the metal conductors [1]. The alternative ways of magnetic writing were proposed: "spin transfer effect" induced by spin-polarized current [2-4], and "magnetoelectric effect", that is the coupling of magnetic and electric subsystems in medium [5]. However the transfer of spin angular momentum requires spin current densities up to $10^7$ A/cm$^2$ that is very close to the strong limitation imposed by electromigration [1]. The considerable progress in the area of magnetoelectric materials has been achieved, especially in thin-film deposition techniques [6]. There were several reports on electric field control of magnetism but it is observed either at low temperatures [7-10] or in artificial media like exchanged coupled bilayer [11] and magnetostrictive/piezoelectric composites [12, 13].

In paper [14] the effect of electric field induced magnetic domain wall displacement in iron garnet films was discovered. The characteristic features of the effect (the dependence of the displacement on the electric polarity and independence of magnetic polarity of the domain) evidenced for magnetoelectric nature of the effect. It provides with the control of magnetization that is realized in single phase material at room temperature by usage electric field only, not implying magnetic field or charge/spin carriers transport.

This paper is focused on the dynamic measurements of the electric field driven domain wall motion that enable us to estimate the effective magnetic field that is characteristic quantity of the effect. Both reversible and irreversible electric field induced micromagnetic structure transformation is demonstrated.

## Experiment

In our experiments we used t used the 9.7-μm-thick epitaxial (BiLu)$_3$(FeGa)$_5$O$_{12}$ ferrite garnet films grown on a Gd$_3$Ga$_5$O$_{12}$ substrate with the (210) crystallographic orientation that has demonstrated the maximum effect [15]. The electric field of high strength was produced by a tip electrode (curvature radius R$_{tip}$ ~ 5 μm) touching the surface of dielectric sample, the magnetooptical technique in Faraday geometry was used to observe the micromagnetic structure

(the experimental details are described elsewhere [14]). For dynamic measurements the high speed photography technique [16] was used: the pulses of electric voltage at the tip (pulse width ~ 300 ns, the rise time ~20 ns) were followed by pulses of laser illumination (duration ~10 ns) to get an image of the structure. Varying the time delay between field and laser pulses enabled us to observe the consecutive positions of domain wall and thus investigate its dynamics.

**Results and discussion**

The most prominent changes were observed at stripe domain heads. In response to the applied electric field domain steadily moves until it reaches the equilibrium position corresponding to the field applied (the consecutive positions of stripe domain head are shown in fig. 1)

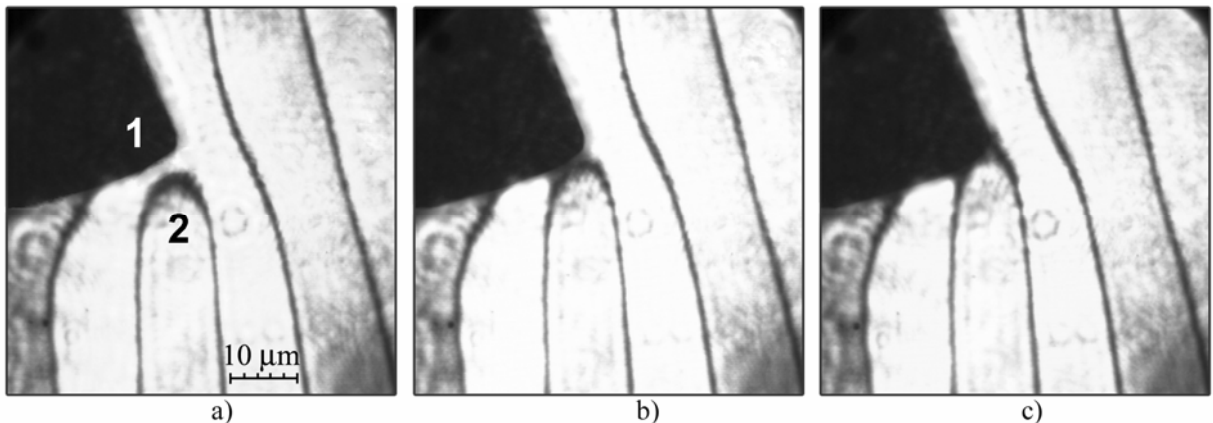

Fig. 1 The stripe domain head moving in magnetoelectric iron garnet film under step-like voltage pulse +400V
a)     the initial configuration: 1 is tip electrode and 2 is the stripe domain head; the consecutive stripe domain head positions: a) before electric pulse (0 ns); b) instantaneous position of the domain wall moving in electric field pulse (75 ns from the start of the pulse); c) at ultimate equilibrium position in electric field (100ns from the pulse start).

Both the values of domain wall velocity and the ultimate displacement of domain wall increase with the value of electric field (fig. 2), but the first value is more adequate characteristic of the effect as it depends on mobility of the wall (that can be easily measured in pulse magnetic field), while the later one is determined by long range magnetostatic forces and their accounting requires complex micromagnetic calculations.

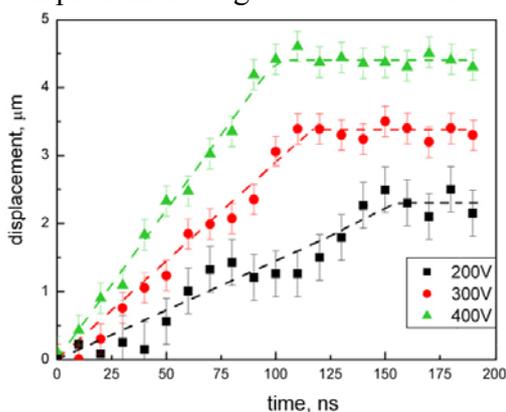

Fig. 2. (Color online) The electric field induced domain wall displacement vs. time. The linear fit of ascending sections of the curves gives the values of the velocities 14±3 m/s, 29±5 m/s, and 44±5 m/s for 200V (black squares), 300V (red circles), and 400V (green triangles) respectively. The large dispersion of experimental points for dynamic dependences in low field (200V) probably is related to domain wall pinning at the defects

The time dependency of the domain wall displacement in 400 V pulse (Fig. 2, green line) enables us to estimate the speed of the wall in the linear region of the dependence as 44±5 m/s.

To compare the velocities achieved in electric field with typical velocities of domain wall in magnetic field we carried out the measurements in magnetic field pulses. The velocity of 50 m/s similar to that one obtained in electric field $E = V/R_{tip} = 800$ kV/cm was achieved in pulse magnetic field about H = 50 Oe.

The largest reversible displacement of the domain wall observed was about 5 μm.

In some cases the electric field caused irreversible displacement of the domain walls of more than 10 μm resulting in dramatic changes of micromagnetic structure (Fig. 3). This experiment could serve as prototypical example of electrical write/magnetic read memory [17] realized in iron garnet media.

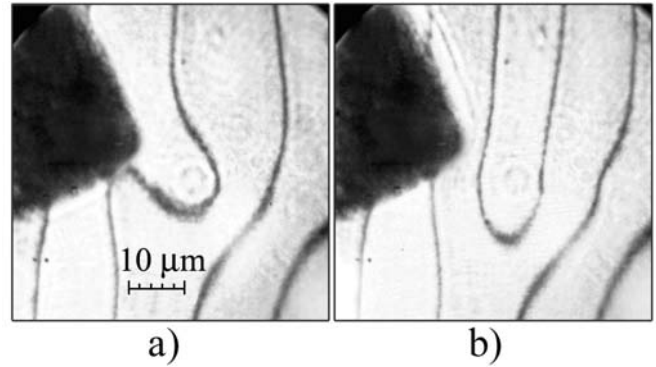

Fig. 3. Irreversible motion of domain wall leading to the transformation of micromagnetic structure. a) the initial position b) the voltage of 500V is applied.

## Conclusion

In conclusion, the electric field driven domain wall motion is implemented in single crystal material at room temperature not implying electric current. The domain wall displacement has reversible character in the range of 1÷5 μm and irreversible one at larger distances. The average domain wall velocity of 50 m/s in 800 kV/cm electric field pulses was achieved that was equivalent to the effect of 50 Oe magnetic field pulses. By further miniaturization of the electrode to nanometric size (the curvature radius of one used in experiment was ~5 μm) the control voltages can be scaled down to several Volts. This effect opens exciting possibilities in the field of spintronics and a route to new scheme of non-volatile memory.